\documentclass[aps,showpacs]{revtex4}

\usepackage{graphics}
\usepackage{epsfig}

\def\ov#1{\overline{#1}}

\def\wt#1{\widetilde{#1}}
\def\vb#1{\mbox{\boldmath$#1$}}
\def\pd#1#2{\frac{\partial #1}{\partial #2}}

\def\wh#1{\widehat{#1}}
\def\bdot{\,\vb{\cdot}\,}
\def\btimes{\,\vb{\times}\,}

\def\bhat{\wh{{\sf b}}}
\def\cal#1{\mathcal{#1}}

\def\exd{{\sf d}}

\newcommand{\bc}{\begin{center}}
\newcommand{\ec}{\end{center}}
\newcommand{\bt}{\begin{tabbing}}
\newcommand{\et}{\end{tabbing}} 
\newcommand{\be}{\begin{eqnarray*}}
\newcommand{\ee}{\end{eqnarray*}}
\newcommand{\bs}{\begin{slide}}
\newcommand{\es}{\end{slide}}

\begin{document}
\begin{flushright}
September 23, 2009 
\par\end{flushright}

\title{Orbit-averaged Guiding-center Fokker-Planck Operator}

\author{A.~J.~Brizard$^{1}$, J.~Decker$^{2}$, Y.~Peysson$^{2}$, and
F.-X.~Duthoit$^{2}$}

\affiliation{$^{1}$Department of Chemistry and Physics, Saint Michael's College,
Colchester, VT 05439, USA \\
 $^{2}$CEA, IRFM, F-13108, Saint-Paul-lez-Durance, France}

\begin{abstract}
A general orbit-averaged guiding-center Fokker-Planck operator suitable
for the numerical analysis of transport processes in axisymmetric
magnetized plasmas is presented. The orbit-averaged guiding-center
operator describes transport processes in a three-dimensional guiding-center
invariant space: the orbit-averaged magnetic-flux invariant $\ov{\psi}$,
the minimum-B pitch-angle coordinate $\xi_{0}$, and the momentum
magnitude $p$. 
\end{abstract}


\maketitle

\section{Introduction}

Transport processes play a dominant role in the long-time behavior
of strongly-magnetized plasmas. In the absence of wave-induced (or
turbulent) transport, the long-time magnetic confinement of plasmas
is based on the small dimensionless parameter $\epsilon_{B}\equiv\rho/L_{B}\ll1$
(the ratio of the characteristic gyroradius $\rho$ and the magnetic
nonuniformity length scale $L_{B}$). For such plasmas, the dimensionless
parameter $\epsilon_{\nu}\equiv L_{B}/\lambda_{\nu}$ ($\lambda_{\nu}$
being the mean-free-path) can be used to describe different classes
of collisional transport processes, such as classical collisional
transport \cite{Braginskii} $(\lambda_{\nu}\ll L_{B})$ and neoclassical
{}``collisionless'' transport \cite{HH_76} $(\lambda_{\nu}\gg L_{B})$.
The quasilinear (wave-induced) transport processes associated with
rf-induced heating and current drive, on the other hand, cause a slow
time evolution of the background plasma distribution (as a result
of one or more wave-particle resonances) with a time-scale ordering
that is quadratic in a small parameter $\epsilon_{w}=|{\bf B}_{w}|/|{\bf B}_{0}|\ll1$
associated with the rf-wave amplitude \cite{Kaufman_72,Brizard_rql_04}.

Because the long-time behavior of the plasma distribution function
for each particle species depends on competing collisional and quasilinear
transport processes, an accurate treatment of both transport processes
in realistic magnetic geometry is a crucial element in determining
the equilibrium and behavior of fusion plasmas. For this purpose,
the use of dynamical-reduction methods (e.g., guiding-center transformation
\cite{RGL_83,Cary_Brizard}) can yield reduced transport operators
(in phase space) that possess attractive numerical properties in addition
to accurately representing collisional \cite{Brizard_2004} and/or
quasilinear \cite{Brizard_rql_04} transport processes of interest.

The purpose of the present paper is to present a brief derivation
of a general orbit-averaged guiding-center Fokker-Planck operator
suitable for numerical studies of transport processes in general axisymmetric
magnetic geometry. This reduced Fokker-Planck operator represents
drag and diffusion processes in a three-dimensional space composed
of guiding-center invariants. In this reduced formulation, the conjugate
orbital angles have either been eliminated from the guiding-center
Fokker-Planck operator by averaging or are absent by axisymmetry.

The remainder of the paper is organized as follows. In Sec.~\ref{sec:gc_axi},
we discuss the guiding-center Hamiltonian dynamics of charged particles
in unperturbed axisymmetric magnetic geometry. In Sec.~\ref{sec:gcFP}, we introduce the
guiding-center Fokker-Planck operator previously derived for general
magnetic geometry and arbitrary guiding-center orbit topology \cite{Brizard_2004}.
In Sec.~\ref{sec:bounce_average}, we first introduce the orbit-averaging operation for standard and non-standard
guiding-center orbits in axisymmetric tokamak geometry. Next, we present the orbit-averaged guiding-center
Fokker-Planck operator and discuss its properties. In Sec.~\ref{sec:bgcFP}, we briefly discuss the derivation of 
the bounce-center Fokker-Planck operator obtained by performing the bounce-center phase-space transformation \cite{Brizard_2000,Brizard_2007}
on the guiding-center Fokker-Planck operator and discuss its connection to the orbit-averaged guiding-center
Fokker-Planck operator derived in the previous section. Lastly, we summarize our work in Sec.~\ref{sec:sum} and discuss its applications.

\section{\label{sec:gc_axi}Guiding-center Dynamics in Axisymmetric Magnetic Geometry}

The existence of the small parameter $\epsilon_{{\rm B}}\ll1$ in
magnetically-confined plasmas forms the basis of the unperturbed guiding-center
dynamical reduction \cite{Cary_Brizard}, in which the fast gyromotion
time scale associated with the gyroangle $\zeta_{{\rm g}}$ (with
$\dot{\zeta_{{\rm g}}}\equiv\epsilon_{{\rm B}}^{-1}\Omega$), with
the gyroaction $J_{{\rm g}}\equiv\mu\, B/\Omega$ acting as its canonically-conjugate
(adiabatic) invariant, is asymptotically decoupled from the parallel
and cross-field motions of a guiding-center particle. The unperturbed
guiding-center dynamics is expressed in terms of the guiding-center
position ${\bf X}$ and the parallel velocity $v_{\|}$, where the
guiding-center velocity 
\begin{equation}
\dot{{\bf X}}\;=\; v_{\|}\;\bhat\;+\;\epsilon_{{\rm B}}\;\frac{\bhat}{m\Omega}\btimes
\left(\mu\;\nabla B\;+\; mv_{\|}^{2}\;\bhat\bdot\nabla\bhat\right)\;\equiv\; v_{\|}\,\bhat\;+\;\epsilon_{{\rm B}}\;{\bf v}_{{\rm B}}
\label{eq:Xdot_gc}
\end{equation}
is decomposed in terms of a parallel velocity $v_{\|}\equiv\bhat\bdot\dot{{\bf X}}$ along a field line and a slower cross-field drift velocity 
${\bf v}_{{\rm B}}$ due to weak magnetic-field nonuniformity. The guiding-center parallel acceleration 
\begin{equation}
\dot{v}_{\|}\;=\;-\;\frac{\mu}{m}\;\left(\bhat\;+\;\epsilon_{{\rm B}}\;\frac{{\bf v}_{{\rm B}}}{v_{\|}}\right)\bdot\nabla B
\label{eq:vpardot_gc}
\end{equation}
guarantees that the guiding-center kinetic energy 
\begin{equation}
{\cal E}\;=\;\frac{m}{2}\; v_{\|}^{2}\;+\;\mu\, B\;\equiv\;\frac{p^{2}}{2m}
\label{eq:E_def}
\end{equation}
is a constant of motion ($\dot{{\cal E}}\equiv0$ for a time-independent magnetic field and in the absence of an electric field). In 
Eqs.~(\ref{eq:Xdot_gc})-(\ref{eq:vpardot_gc}), the guiding-center magnetic-moment invariant $\mu$ and the coordinates $({\bf X},v_{\|})$ in the reduced four-dimensional guiding-center phase space are expressed as asymptotic expansions in powers of the small parameter $\epsilon_{{\rm B}}$, where first-order corrections explicitly take into account magnetic-field nonuniformity \cite{RGL_83}.

We note that, while a nonrelativistic guiding-center formulation is considered here, its generalization to a relativistic formulation (appropriate
for fast electrons) can easily be accommodated \cite{Brizard_rgc}. The kinetic energy (\ref{eq:E_def}) is thus replaced with ${\cal E}=(\gamma-1)\, 
mc^{2}$, the parallel velocity $v_{\|}$ is replaced with the relativistic parallel momentum $p_{\|}=\gamma\, mv_{\|}$, and the magnetic moment
$\mu$ is replaced with the relativistic magnetic moment $\mu=|{\bf p}_{\bot}|^{2}/2mB$, where the relativistic factor is $\gamma=(1+2\,\mu B/mc^{2}+
p_{\|}^{2}/m^{2}c^{2})^{1/2}$. Additional details on relativistic guiding-center dynamics in axisymmetric magnetic geometry can be found in 
Ref.~\cite{Cooper}.

\subsection{Axisymmetric Magnetic geometry}

The general axisymmetric magnetic field considered in the present
paper is expressed in terms of three equivalent representations \cite{White_2008}:
\begin{equation}
{\bf B}\;=\;\left\{ \begin{array}{l}
\nabla\phi\btimes\nabla\psi\;+\; q(\psi)\;\nabla\psi\btimes\nabla\theta\\
\\B_{\phi}\,\nabla\phi\;+\; B_{\theta}\,\nabla\theta\;+\; B_{\psi}\,\nabla\psi\\
\\B^{\phi}\;\partial{\bf X}/\partial\phi\;+\; B^{\theta}\;\partial{\bf X}/\partial\theta\end{array}\right.
\label{eq:B_axisym}
\end{equation}
 where the two-covariant, covariant, and contravariant representations
(from top to bottom, respectively) are expressed in terms of the (poloidal)
magnetic flux $\psi$, which satisfies the condition ${\bf B}\bdot\nabla\psi=0$
(i.e., magnetic field lines lie entirely on a constant-$\psi$ surface),
and the poloidal and toroidal angles $\theta$ and $\phi$. The toroidal
and poloidal components of the magnetic field (\ref{eq:B_axisym})
are $B_{{\rm tor}}=B_{\phi}/R=B^{\phi}\, R$, where $R\equiv|\partial{\bf X}/\partial\phi|=|\nabla\phi|^{-1}$,
and $B_{{\rm pol}}=|\nabla\psi|/R=B^{\theta}\,|\partial{\bf X}/\partial\theta|$.
The safety factor $q(\psi)$ appearing in the two-covariant representation
is defined as \cite{footnote_1} 
\begin{equation}
q(\psi)\;\equiv\;\frac{{\bf B}\bdot\nabla\phi}{{\bf B}\bdot\nabla\theta}\;=\;\frac{B^{\phi}}{B^{\theta}}.
\label{eq:q_def}
\end{equation}
We note that, because of the magnetic-flux condition ${\bf B}\bdot\nabla\psi=0$, the covariant component $B_{\psi}\equiv-\, B_{\theta}\,(\nabla\theta\bdot\nabla\psi)/|\nabla\psi|^{2}$ in Eq.~(\ref{eq:B_axisym}) vanishes only if the coordinates $\psi$
and $\theta$ are orthogonal. The spatial Jacobian ${\cal V}$ associated
with the coordinates $(\psi,\theta,\phi)$ is 
\begin{equation}
{\cal V}\;\equiv\;(\nabla\psi\btimes\nabla\theta\bdot\nabla\phi)^{-1}\;=\;({\bf B}\bdot\nabla\theta)^{-1}\;=\;(B^{\theta})^{-1},
\label{eq:Jac_space}
\end{equation}
 where $B^{\theta}$ is assumed to be positive. Lastly, the infinitesimal
length element along a magnetic field line is 
\begin{equation}
ds\;\equiv\;\frac{B}{B^{\theta}}\; d\theta.
\label{eq:ds_def}
\end{equation}
 Note that, according to the standard axisymmetric tokamak ordering
(with $B\simeq B_{{\rm tor}}$), the ratio $B/B^{\theta}$ can also
be expressed as $B/B^{\theta}=B/[B_{{\rm tor}}/(qR)]\simeq q\, R$,
so that we recover the standard approximation $ds\simeq qR\, d\theta$.

\subsection{Guiding-center Motion in Axisymmetric Magnetic geometry}

The guiding-center motion in arbitrary magnetic geometry described
by Eqs.~(\ref{eq:Xdot_gc})-(\ref{eq:vpardot_gc}) possesses two
constants of the motion: the total guiding-center energy (\ref{eq:E_def})
and the guiding-center magnetic moment $\mu$. Guiding-center motion
in general axisymmetric magnetic geometry (\ref{eq:B_axisym}) is
also characterized, according to Noether's theorem, by a third constant
of the motion: the toroidal canonical guiding-center momentum 
\begin{equation}
P_{\phi}\;\equiv\;\pd{\bf X}{\phi}\bdot\left(\frac{e}{c}\,{\bf A}\;+\; m\, v_{\|}\;\bhat\right)\;=\;-\,\frac{e}{c}\left(\psi\;-\frac{}{}\rho_{\|}\; B_{\phi}\right).
\label{eq:p_phi}
\end{equation}
 Here, the vector potential ${\bf A}=-\,\psi\,\nabla\phi+\psi_{{\rm tor}}(\psi)\,\nabla\theta$
(with $q\equiv d\psi_{{\rm tor}}/d\psi$ defined in terms of the toroidal
magnetic flux $\psi_{{\rm tor}}$) was obtained from the two-covariant
representation in Eq.~(\ref{eq:B_axisym}), where the coordinates
$(\psi,\theta,\phi)$ now describe the guiding-center position ${\bf X}$,
$\rho_{\|}\equiv v_{\|}/\Omega$ denotes the parallel gyroradius,
and $B_{\phi}\equiv{\bf B}\bdot\partial{\bf X}/\partial\phi$ denotes
the covariant component of the axisymmetric magnetic field.

The projection of the two-dimensional drift surface \cite{HH_76}
\begin{equation}
\psi\;-\;\rho_{\|}\; B_{\phi}\;\equiv\;\ov{\psi}
\label{eq:psi_drift}
\end{equation}
 onto the poloidal plane $(X,Z)$ generates a closed curve $\psi=\wt{\psi}(\theta)$
parameterized by the poloidal angle $\theta$ and labeled by the guiding-center
invariants $(\ov{\psi},{\cal E},\mu)$: 
\begin{equation}
\wt{\psi}(\theta,\sigma;\ov{\psi},{\cal E},\mu)\;\equiv\;\ov{\psi}\;+\;\delta\psi(\theta,\sigma;\ov{\psi},{\cal E},\mu),
\label{eq:psi_theta}
\end{equation}
 where the {}``bounce-radius'' $\delta\psi=\wt{\psi}-\ov{\psi}$
represents the departure of the drift surface (\ref{eq:psi_theta}),
labeled by $\ov{\psi}$, from a magnetic surface $\psi$ and $\sigma=\pm\,1$
denotes the sign of $v_{\|}$. A guiding-center orbit ${\cal O}$
is obtained either by integrating the guiding-center equations of
motion 
\begin{equation}
\left. \begin{array}{rcl}
\dot{\psi} & \equiv & \dot{{\bf X}}\bdot\nabla\psi \;=\; v_{\|}\,\bhat\bdot\nabla\psi \;+\; \epsilon_{\rm B}\,{\bf v}_{{\rm B}}\bdot\nabla\psi
\;\equiv\; \epsilon_{\rm B}\;\dot{\psi}_{\rm B} \\
 &  & \\
\dot{\theta} & \equiv & \dot{{\bf X}}\bdot\nabla\theta \;=\; v_{\|}\,\bhat\bdot\nabla\theta \;+\; \epsilon_{\rm B}\,{\bf v}_{{\rm B}}\bdot\nabla\theta
\;\equiv\;  v_{\|}\,B^{\theta}/B \;+\; \epsilon_{\rm B}\;\dot{\theta}_{\rm B}
\end{array} \right\}
\label{eq:psitheta_dot}
\end{equation}
for a given set of guiding-center invariants $(\ov{\psi},{\cal E},\mu)$ or generating the orbit directly by the constant-of-motion method \cite{Rome_Peng}.
For each {\it generic} set of guiding-center invariants $(\ov{\psi},{\cal E},\mu)$, there corresponds a unique guiding-center orbit ${\cal O}$, which is either a \textit{trapped-particle} orbit (if the bounce-radius $\delta\psi$ vanishes along the orbit), or a \textit{passing-particle} orbit (if the bounce-radius $\delta\psi$ does not vanish). Non-generic orbits \cite{Review_orbits} include the {\it stagnation} orbits (where $\dot{\theta} = 0$ or
$v_{\|}\,B^{\theta}/B = -\,\epsilon_{\rm B}\;\dot{\theta}_{\rm B} \neq 0$) and the {\it pinch} orbits (barely-trapped orbits in the zero-banana-width limit).

On each flux-surface $\psi$, the magnetic field amplitude is assumed
to vary monotonically between a minimum value $B_{0}(\psi)$ and a
maximum value $B_{1}(\psi)$, located at the poloidal locations $\theta_{0}(\psi)$
and $\theta_{1}(\psi)$, respectively. The turning points $(\psi_{{\rm b}},\theta_{{\rm b}}^{\pm})$
of a trapped-particle orbits $(\ov{\psi},{\cal E},\mu)$ are located
on the drift-surface label $\psi_{{\rm b}}=\ov{\psi}$, where the
magnetic field reaches its maximum value along the orbit and $\delta\psi(\theta_{{\rm b}}^{\pm},\pm1;\psi_{{\rm b}},{\cal E},\mu)\equiv0$.
Consequently, the X point separating trapped-particle and passing-particle orbits
parameterized by the drift-surface label $\ov{\psi}$ is located at
the position $[\ov{\psi},\theta_{1}(\ov{\psi})]$.

While the magnetic moment $\mu$ is an important invariant for guiding-center
dynamics, the pitch-angle coordinate 
\begin{equation}
\xi(\wt{\psi};{\cal E},\mu)\;\equiv\; v_{\|}/v\;=\;\sigma\;\sqrt{1\;-\;\mu B(\wt{\psi})/{\cal E}}.
\label{eq:xi_def}
\end{equation}
is better suited in describing the transition between trapped-particle
and passing-particle orbits. In order to convert the pitch-angle coordinate
into a suitable guiding-center invariant, we replace the guiding-center
magnetic-moment invariant $\mu$ with {[}16{]} 
\begin{equation}
\xi_{0}(\ov{\psi},{\cal E},\mu)\;\equiv\;\sqrt{1-\mu B_{0}(\ov{\psi})/{\cal E}}.\label{eq:xi0_def}\end{equation}
 The physical interpretation of the pitch-angle invariant $\xi_{0}$
can be given in terms of its connection with the bounce-action invariant
$J_{{\rm b}}$. With this definition, the trapped-passing separatrix
given by the relation,
\[ 1\;-\;(1-\xi_{0}^{2})\; B_{1}(\ov{\psi})/B_{0}(\ov{\psi})=0 \]
which is an even function of $\xi_{0}$ and does not depend upon the energy
${\cal E}$ (important for numerical applications).

The guiding-center Jacobian in coordinates $({\bf X},p,\xi,\zeta_{{\rm g}})$
is ${\cal J}_{{\rm gc}}\equiv p^{2}$, while in terms of the coordinates
$({\bf X},p,\xi_{0},\zeta_{{\rm g}})$, it is ${\cal J}_{{\rm gc}0}\equiv p^{2}\,|\partial\xi/\partial\xi_{0}|=p^{2}\;\Psi\,\xi_{0}/|\xi|$,
where we used the definition $\Psi(\wt{\psi},\theta)\equiv B(\wt{\psi},\theta)/B_{0}(\ov{\psi})$
with the relation
\begin{equation}
|\xi|\;=\;\sqrt{1\;-\;\Psi\;\left(1-\xi_{0}^{2}\right)}.
\label{eq:Psi_def}
\end{equation}
Next, the spatial Jacobian in coordinates ${\bf X}\equiv(\ov{\psi},\theta,\phi)$
is ${\cal V}=[B^{\theta}(\wt{\psi})]^{-1}$, so that the total guiding-center
Jacobian in coordinates $(\ov{\psi},\theta,\phi;\, p,\xi_{0},\zeta_{{\rm g}})$
is \begin{equation}
{\cal J}\;\equiv\;{\cal V}\;{\cal J}_{{\rm gc}0}\;=\; p^{2}\;\frac{\Psi}{B^{\theta}}\,\frac{\xi_{0}}{|\xi|}\;=\;\frac{p^{2}}{B_{0}}\;
\left(\frac{B\,\xi_{0}}{B^{\theta}\,|\xi|}\right).
\label{eq:Jac_def}
\end{equation}
In the next section, we express a general guiding-center Fokker-Planck operator in terms of the guiding-center invariants $(\ov{\psi},p,\xi_{0})$ in a form that will be suitable for orbit averaging in Sec.~\ref{sec:bounce_average}.

\section{\label{sec:gcFP}General Guiding-center Fokker-Planck Equation}

The Fokker-Planck equation \cite{HH_76,Kaufman_72,FP_review} forms
a paradigm for the investigation of classical, neoclassical, and quasilinear
transport processes in plasmas. When written in terms of the test-particle
phase-space coordinates $({\bf x},{\bf p})$, the general Fokker-Planck
operator is expressed as a local partial-differential operator in
momentum space \cite{HH_76}: 
\begin{equation}
{\cal C}[f]({\bf x},{\bf p})\;=\;-\;\pd{}{\bf p}\bdot\left({\bf K}\; f({\bf x},{\bf p})\;-\;{\sf D}\bdot\pd{f({\bf x},{\bf p})}{\bf p}\right),
\label{eq:FP_def}
\end{equation}
 where the Fokker-Planck momentum-friction vector ${\bf K}$ and the
momentum-diffusion tensor ${\sf D}$ are functions of $({\bf x},{\bf p})$.
When representing collisional transport processes, for example, these
Fokker-Planck coefficients ${\bf K}\equiv\sum^{\prime}\,{\bf K}[f^{\prime}]$
and ${\sf D}\equiv\sum^{\prime}\,{\sf D}[f^{\prime}]$ are expressed
as integral operators acting on the field-particle distribution $f^{\prime}$
(where the field-particle species may coincide with the test-particle
species). Hence, the Fokker-Planck operator (\ref{eq:FP_def}) may
either be a linear, bilinear, or nonlinear operator acting $f$, depending
on the type of transport problem one wishes to investigate.

While collisions and wave-particle interactions take place locally
in particle phase space ${\bf z}=({\bf x},{\bf p})$, the transformation
to reduced phase-space coordinates $\ov{\bf z}$ (e.g., guiding-center
coordinates) will generically introduce transport coefficients in
the full reduced phase space. This implies that the Fokker-Planck
coefficients $(K^{i},D^{ij})$ in three-dimensional particle-momentum
space are replaced with six-dimensional reduced phase-space Fokker-Planck
coefficients $(\ov{K}^{\alpha},\ov{D}^{\alpha\beta})$, defined as
\begin{equation}
\ov{K}^{\alpha}(\ov{\bf z})\;\equiv\;\left[K^{i}({\bf z})\;\pd{\ov{z}^{\alpha}({\bf z})}{p^{i}}\right]_{{\bf z}={\bf z}(\ov{\bf z})}
\;\;{\rm and}\;\;\ov{D}^{\alpha\beta}(\ov{\bf z})\;\equiv\;\left[\pd{\ov{z}^{\alpha}({\bf z})}{p^{i}}\; 
D^{ij}({\bf z})\;\pd{\ov{z}^{\beta}({\bf z})}{p^{j}}\right]_{{\bf z}={\bf z}(\ov{\bf z})},
\label{eq:ovKD_def}
\end{equation}
 and the Fokker-Planck operator (\ref{eq:FP_def}) transforms to 
\begin{equation}
\ov{\cal C}[\ov{f}](\ov{\bf z})\;=\;-\;\frac{1}{\ov{\cal J}}\pd{}{\ov{z}^{\alpha}}\left[\;\ov{\cal J}\left(\ov{K}^{\alpha}\;\ov{f}\;-\;\ov{D}^{\alpha\beta}\;\pd{\ov{f}}{\ov{z}^{\beta}}\right)\;\right],
\label{eq:ovFP_def}
\end{equation}
 where $\ov{f}(\ov{\bf z})\equiv f({\bf z})$ is the particle distribution
expressed in terms of the reduced phase-space coordinates and $\ov{\cal J}$
is the Jacobian for the transformation ${\bf z}\rightarrow\ov{\bf z}$.
The most important aspect of the transformation to Eq.~(\ref{eq:ovFP_def})
involves the choice of the new phase-space coordinates $\ov{\bf z}$.
Note that the definitions (\ref{eq:ovKD_def}) for the reduced phase-space
Fokker-Planck coefficients require that the transformation ${\bf z}\rightarrow\ov{\bf z}({\bf z})$
and its inverse $\ov{\bf z}\rightarrow{\bf z}(\ov{\bf z})$ must be
known up to the desired order in magnetic-field nonuniformity.

One possible choice is to adopt a canonical action-angle formulation
\cite{Bernstein_Molvig}, where $\ov{\bf z}=(\ov{\bf J},\ov{\vb{\theta}})$
includes the three-dimensional action invariants $\ov{\bf J}$ for
magnetically-confined particles and their canonically-conjugate angles
$\ov{\vb{\theta}}$ (which are ignorable coordinates by construction
$\partial\ov{f}/\partial\ov{\vb{\theta}}\equiv0$). While the three-dimensional
action-space Fokker-Planck operator derived by Bernstein and Molvig
\cite{Bernstein_Molvig} formally describes classical and neoclassical
transport processes in axisymmetric magnetic geometry, it is not suitable
for numerical implementation since some of the action coordinates
(e.g., the bounce action $J_{{\rm b}}$) are not \textit{local} coordinates.
Hence, the action-space Fokker-Planck operator does not describe local
transport processes, which makes the transport analysis of its results
difficult to interpret.

Another choice is to adopt \textit{local} noncanonical guiding-center
coordinates \cite{ZOC_93,Brizard_2004} leading to the construction
of a reduced guiding-center Fokker-Planck operator. In a weakly-nonuniform
plasma with a strong axisymmetric magnetic field, for example, the
reduced guiding-center Fokker-Planck equation describes transport
processes in a four-dimensional space: the poloidal-flux and poloidal-angle coordinates
$(\psi,\theta)$ in physical space, and the energy and magnetic moment
coordinates $({\cal E},\mu)$ in velocity space, which are invariants
for the guiding-center motion. Here, the reduction from six to four
dimensions is associated with the fact that the guiding-center plasma
distribution is independent of the gyroangle $\zeta_{{\rm g}}$ (by
definition) and the toroidal angle $\phi$ (by axisymmetry), so that
transport along these ignorable angles is irrelevant. While Zaitsev
\textit{et al.} \cite{ZOC_93} considered the lowest-order definitions
(in magnetic-field nonuniformity) for their choice of guiding-center
coordinates, Brizard \cite{Brizard_2004} considered first-order corrections
($\epsilon_{\rm B} = \rho/L_{{\rm B}}\ll 1$) as well. In both works, classical
collisional transport appears in Eq.~(\ref{eq:ovKD_def}) as a result
of the lowest-order relation $\partial{\bf X}/\partial{\bf p}=-\,\bhat\btimes{\bf I}/m\Omega$
for the guiding-center position ${\bf X}({\bf x},{\bf p})\equiv{\bf x}-\vb{\rho}_{0}({\bf x},{\bf p})$,
where $\vb{\rho}_{0}\equiv\bhat\btimes{\bf p}/m\Omega$ represents
the lowest-order gyroradius. The higher-order guiding-center corrections
kept by Brizard \cite{Brizard_2004} are consistent with the low-collisionality
regime, where the mean-free-path $\lambda_{\nu}>L_{{\rm B}}$ is longer
than the magnetic-nonuniformity length scale $L_{{\rm B}}$.

\subsection{Guiding-center Fokker-Planck Operator}

The derivation of a reduced Fokker-Planck operator that is suitable
for numerical implementation must begin with finding local invariant
coordinates that also allow relative computational simplicity for
realistic magnetic geometries. In the present work, the exact invariants
associated with the time-independent axisymmetric magnetic geometry
are the kinetic momentum $p=\sqrt{2m\,{\cal E}}$ and the drift-surface
label $\ov{\psi}=\psi_{{\rm b}}$ (for trapped particles) or $\psi_{{\rm t}}$
(for passing particles). As our third invariant coordinate, we use
the pitch-angle invariant $\xi_{0}$, defined in Eq.~(\ref{eq:xi0_def}),
which allows an explicit representation of the trapping and detrapping
transport processes.

Once a set of invariant guiding-center coordinates $I^{a} = (\ov{\psi}, p, \xi_{0})$ is chosen
(not necessarily action coordinates), the transformation of the Fokker-Planck operator (\ref{eq:FP_def}) can be greatly simplified by writing it
in Poisson-bracket form as 
\begin{equation}
{\cal C}[f]\;=\;\;\equiv\;-\;\left\{ x^{i},\;\left(K^{i}\; f\;-\; D^{ij}\frac{}{}\left\{ x^{j},\; f\right\} \right)\right\},
\label{eq:FP_PB}
\end{equation}
 where the noncanonical Poisson bracket $\{\;,\;\}$ is used to replace
momentum partial derivatives $\partial g/\partial p_{i}\equiv\{x^{i},g\}$.
The significant computational advantage of this Poisson-bracket formulation
(\ref{eq:FP_PB}) is based on the fact that Poisson brackets transform
naturally under the type of phase-space transformations considered
in the present work (i.e., those generated by Lie-transform methods).

The general guiding-center Fokker-Planck operator was derived by Brizard
\cite{Brizard_2004} as a result of the guiding-center dynamical reduction
of the Fokker-Planck operator (\ref{eq:FP_def}): 
\begin{equation}
{\cal C}_{{\rm gc}}[F]\;\equiv\;\left\langle {\sf T}_{{\rm gc}}^{-1}{\cal C}\left[{\sf T}_{{\rm gc}}\frac{}{}F\right]\right\rangle _{{\rm g}}\;=\;-\;\left\langle \left\{ X^{i}+\rho_{\epsilon}^{i},\;\left({\sf T}_{{\rm gc}}^{-1}K^{i}\; F\;-\;{\sf T}_{{\rm gc}}^{-1}D^{ij}\;\left\{ X^{j}+\rho_{\epsilon}^{j},\; F\right\} _{{\rm gc}}\right)\right\} _{{\rm gc}}\right\rangle _{{\rm g}},
\label{eq:gcFP_operator}
\end{equation}
 where ${\sf T}_{{\rm gc}}$ and ${\sf T}_{{\rm gc}}^{-1}$ denote
the pull-back and push-forward (Lie-transform) operators associated
with the guiding-center transformation for a nonuniform magnetic field,
$\langle\cdots\rangle_{{\rm g}}$ denotes an average with respect
to the guiding-center gyroangle $\zeta_{{\rm g}}$, and $\{\;,\;\}_{{\rm gc}}$
denotes the guiding-center Poisson bracket. In Eq.~(\ref{eq:gcFP_operator}),
the generalized gyroradius vector $\vb{\rho}_{\epsilon}$ contains
first-order corrections associated with magnetic-field nonuniformity.

When we use the phase-space-divergence property of Poisson bracket
\[
\{F,\; G\}_{{\rm gc}}\;\equiv\;\frac{1}{{\cal J}}\;\pd{}{Z^{\alpha}}\left({\cal J}\;\left\{ F,\frac{}{}Z^{\alpha}\right\} _{{\rm gc}}\; G\right),\]
 where $F$ and $G$ are two arbitrary guiding-center phase-space
functions, we can write the divergence form of the guiding-center
Fokker-Planck operator (\ref{eq:gcFP_operator}): 
\begin{equation}
{\cal C}_{{\rm gc}}[F]\;\equiv\;-\;\frac{1}{{\cal J}}\pd{}{Z^{\alpha}}\left[\;{\cal J}\left({\cal K}_{{\rm gc}}^{\alpha}\; F\;-\;{\cal D}_{{\rm gc}}^{\alpha\beta}\;\pd{F}{Z^{\beta}}\right)\;\right],
\label{eq:gcFP_def}
\end{equation}
 where ${\cal J}$ is the total Jacobian (\ref{eq:Jac_def}). The
guiding-center Fokker-Planck friction components 
\begin{equation}
{\cal K}_{{\rm gc}}^{\alpha}\;\equiv\;\left\langle {\sf T}_{{\rm gc}}^{-1}{\bf K}\bdot\vb{\Delta}^{\alpha}\right\rangle _{{\rm g}}
\label{eq:Kgc}
\end{equation}
 and the guiding-center Fokker-Planck diffusion components 
\begin{equation}
{\cal D}_{{\rm gc}}^{\alpha\beta}\;\equiv\;\left\langle (\vb{\Delta}^{\alpha})^{\top}\bdot{\sf T}_{{\rm gc}}^{-1}{\sf D}\bdot\vb{\Delta}^{\beta}\right\rangle _{{\rm g}},
\label{eq:Dgc}
\end{equation}
 are expressed in terms of the guiding-center push-forward expressions
${\sf T}_{{\rm gc}}^{-1}{\bf K}$ and ${\sf T}_{{\rm gc}}^{-1}{\sf D}$
of the particle momentum-space Fokker-Planck friction vector ${\bf K}$
and diffusion tensor ${\sf D}$, and the vector-valued \textit{projection}
coefficients 
\begin{equation}
\vb{\Delta}^{\alpha}\;\equiv\;\left\{ {\bf X}+\vb{\rho}_{\epsilon},\; Z^{\alpha}\right\} _{{\rm gc}}\;=\;\vb{\Delta}^{{\bf X}}\bdot\nabla Z^{\alpha}\;+\;\vb{\Delta}^{{\cal E}}\;\pd{Z^{\alpha}}{\cal E}\;+\;\vb{\Delta}^{\mu}\;\pd{Z^{\alpha}}{\mu}.
\label{eq:Delta_alpha}
\end{equation}
 Here, the guiding-center vector-valued projection coefficients in
$({\bf X},{\cal E},\mu)$ space \cite{Brizard_2004} 
\begin{equation}
\left.\begin{array}{rcl}
\vb{\Delta}^{{\bf X}} & = & \bhat\btimes{\bf I}/(m\Omega)\;=\;-\;(\vb{\Delta}^{{\bf X}})^{\top}\\
\vb{\Delta}^{{\cal E}} & = & {\bf p}_{\epsilon}/m\\
\vb{\Delta}^{\mu} & = & (\Omega/B)\;\partial\vb{\rho}_{\epsilon}/\partial\zeta_{{\rm g}}\end{array}\right\} 
\label{eq:Delta_gc}
\end{equation}
 are constructed from the guiding-center Poisson-bracket expressions
involving the guiding-center push-forward of the particle position
${\sf T}_{{\rm gc}}^{-1}{\bf x}\equiv{\bf X}+\vb{\rho}_{\epsilon}$,
where the generalized gyroradius $\vb{\rho}_{\epsilon}\equiv{\sf T}_{{\rm gc}}^{-1}\vb{\rho}$
and the generalized guiding-center momentum 
\begin{equation}
{\bf p}_{\epsilon}\;\equiv\;{\sf T}_{{\rm gc}}^{-1}{\bf p}\;=\; m\,\left(\frac{d_{{\rm gc}}{\bf X}}{dt}\;+\;\frac{d_{{\rm gc}}\vb{\rho}_{\epsilon}}{dt}\right)
\label{eq:p_epsilon}
\end{equation}
 contain first-order corrections associated with the guiding-center
transformation. Note that, while the projection coefficients $\vb{\Delta}^{{\cal E}}$
and $\vb{\Delta}^{\mu}$ in Eq.~(\ref{eq:Delta_gc}) retain first-order
corrections in magnetic-field nonuniformity, we have omitted first-order
corrections in $\vb{\Delta}^{{\bf X}}$ since they yield second-order
contributions in the final expressions for Eq.~(\ref{eq:gcFP_def}).

\subsection{Guiding-center Projection Vectors}

Using the definitions for the invariant-space coordinates $I^{a}=(\ov{\psi},p,\xi_{0})$, the guiding-center vector-valued projection
coefficients $\vb{\Delta}^{a}=(\vb{\Delta}^{\psi},\vb{\Delta}^{p},\vb{\Delta}^{\xi_{0}})$ are expressed as \cite{Brizard_2004} 
\begin{eqnarray}
\vb{\Delta}^{\ov{\psi}} & \equiv & \vb{\Delta}^{{\bf X}}\bdot\nabla\ov{\psi}\;=\;\frac{\bhat}{m\Omega}\btimes\nabla\ov{\psi},
\label{eq:Delta_psi}\\
\vb{\Delta}^{p} & \equiv & \vb{\Delta}^{{\cal E}}\;\pd{p}{\cal E}\;=\;\frac{{\bf p}_{\epsilon}}{|{\bf p}_{\epsilon}|},
\label{eq:Delta_p}\\
\vb{\Delta}^{\xi_{0}} & \equiv & \vb{\Delta}^{{\bf X}}\bdot\nabla\xi_{0}\;+\;\vb{\Delta}^{{\cal E}}\;\pd{\xi_{0}}{\cal E}\;+\;\vb{\Delta}^{\mu}\;
\pd{\xi_{0}}{\mu}\;=\;\left(\frac{1-\xi_{0}^{2}}{2\;\xi_{0}}\right)\left[\left(\frac{{\bf p}_{\epsilon}}{m\,{\cal E}}\;-\;\frac{\Omega}{\mu B}\,
\pd{\vb{\rho}_{\epsilon}}{\zeta_{{\rm g}}}\right)\;-\;\vb{\Delta}^{\ov{\psi}}\;\frac{d\ln B_{0}}{d\ov{\psi}}\right].
\label{eq:Delta_xi0}
\end{eqnarray}
 We now discuss the physical nature of the projection vectors (\ref{eq:Delta_psi})-(\ref{eq:Delta_xi0}).

\subsubsection{Guiding-center Radial Projection}

Using the magnetic representation (\ref{eq:B_axisym}), the guiding-center
radial projection vector (\ref{eq:Delta_psi}) becomes 
\begin{equation}
\vb{\Delta}^{\ov{\psi}}\;=\;\frac{\bhat}{m\Omega}\btimes\nabla\ov{\psi}\;=\;-\;\frac{c}{e\, B^{2}}\,\left[|\nabla\ov{\psi}|^{2}\;\nabla\phi\;-\frac{}{}q\;\left(\nabla\ov{\psi}\btimes\nabla\theta\right)\btimes\nabla\ov{\psi}\right]\;\equiv\;-\;\frac{c}{e}\,\frac{|\nabla\ov{\psi}|^{2}}{B^{2}}\;\nabla_{\psi}\chi,
\label{eq:Deltapsi_exp}
\end{equation}
 where we introduced the angle-like coordinate $\chi\equiv\phi\,-\, q(\ov{\psi})\,\theta$,
such that the two-covariant representation of the axisymmetric magnetic
field (\ref{eq:B_axisym}) becomes ${\bf B}\equiv\nabla\chi\btimes\nabla\ov{\psi}$.
Here, $\nabla_{\psi}\theta\equiv(\nabla\ov{\psi}\btimes\nabla\theta)\btimes\nabla\ov{\psi}/|\nabla\ov{\psi}|^{2}$
denotes the projection of the gradient of $\theta$ that is perpendicular
to $\nabla\ov{\psi}$, while $\nabla_{\psi}\phi\equiv\nabla\phi$
(since $\nabla\phi\bdot\nabla\ov{\psi}\equiv0$). Hence, the guiding-center
radial projection vector (\ref{eq:Deltapsi_exp}) projects momentum-space
transport processes onto a magnetic-flux surface in a direction that
is locally perpendicular to $\bhat$.

The guiding-center radial projection vector $\vb{\Delta}^{\ov{\psi}}$,
whose magnitude is \[
|\vb{\Delta}^{\ov{\psi}}|\;\equiv\;\frac{|\nabla\ov{\psi}|}{m\Omega}\;=\;\frac{cB_{{\rm pol}}}{eB}\; R,\]
 therefore generates the local projection of the guiding-center push-forwards
of the momentum-space Fokker-Planck friction 
\begin{equation}
{\cal K}_{{\rm gc}}^{\ov{\psi}}\;=\;\left\langle {\sf T}_{{\rm gc}}^{-1}{\bf K}\right\rangle _{{\rm g}}\bdot\frac{\bhat}{m\Omega}\btimes\nabla\ov{\psi},
\label{eq:Kgc_psi}
\end{equation}
 and the momentum-space Fokker-Planck diffusion 
\begin{eqnarray}
{\cal D}_{{\rm gc}}^{a\,\ov{\psi}} & = & \left\langle \vb{\Delta}^{a}\bdot{\sf T}_{{\rm gc}}^{-1}{\sf D}\right\rangle _{{\rm g}}\bdot\frac{\bhat}{m\Omega}\btimes\nabla\ov{\psi},
\label{eq:Dgc_alpha_psi}\\
{\cal D}_{{\rm gc}}^{\ov{\psi}\ov{\psi}} & = & \frac{\bhat}{m\Omega}\btimes\nabla\ov{\psi}\bdot\left\langle {\sf T}_{{\rm gc}}^{-1}{\sf D}
\right\rangle _{{\rm g}}\bdot\frac{\bhat}{m\Omega}\btimes\nabla\ov{\psi},
\label{eq:Dgc_psipsi}
\end{eqnarray}
 where $a\neq\ov{\psi}$ in Eq.~(\ref{eq:Dgc_alpha_psi}). These
projections only take into account transport processes occuring within
a magnetic-flux surface. The connection between radial transport and
toroidal and/or poloidal rotation is therefore naturally contained
within this projection.

\subsubsection{Guiding-center Energy Projection}

The guiding-center energy projection vector (\ref{eq:Delta_p}) is
expressed as 
\begin{equation}
\vb{\Delta}^{p}\;\equiv\;\frac{{\bf p}_{\epsilon}}{|{\bf p}_{\epsilon}|}\;=\;\wh{\sf p}\;+\;\epsilon\;\vb{\Delta}_{1}^{p}\;+\;\cdots,
\label{eq:Deltap_exp}
\end{equation}
 where $\vb{\Delta}_{1}^{p}$ denotes the first-order correction of
$\vb{\Delta}^{p}$. It can be explicitly expressed in terms of the
gyroangle-independent first-order gyroradius $\ov{\vb{\rho}}_{{\rm gc}}\equiv(\bhat/\Omega)\btimes{\bf v}_{{\rm B}}$
and $\vb{\rho}_{\epsilon}\equiv\vb{\rho}_{0}+\epsilon\;\vb{\rho}_{1}$
as \cite{Brizard_2004} 
\begin{equation}
\vb{\Delta}_{1}^{p}\;=\;\frac{1}{\rho}\;\left(\ov{\vb{\rho}}_{{\rm gc}}\btimes\bhat\;+\;\pd{\vb{\rho}_{1}}{\zeta_{{\rm g}}}\right)
\;+\;\xi\;\bhat\bdot\nabla^{*}\vb{\rho}_{0}\;\equiv\;\Delta_{1\xi}^{p}\;\pd{\wh{\sf p}}{\xi}\;+\;\Delta_{1\zeta}^{p}\;
\pd{\wh{\sf p}}{\zeta_{{\rm g}}}.
\label{eq:Deltap_1}
\end{equation}
 where $\rho\equiv p/(m\Omega)$ is the gyroradius magnitude for a
deeply-trapped particle (i.e., for $\xi_{0}=0$) and 
\[ \frac{1}{\rho}\;\ov{\vb{\rho}}_{{\rm gc}}\btimes\bhat\;=\;\rho\;\bhat\btimes\left[\frac{1}{2}\,\left(1-\xi^{2}\right)\;\nabla\ln B\;+\;\xi^{2}\;\bhat\bdot\nabla\bhat\right]. \]
 It can be shown explicitly, however, that $\vb{\Delta}_{1}^{p}$
does not have any component directed along the zeroth-order unit vector
$\wh{\sf p}$ (because the guiding-center kinetic energy is identical
to the particle kinetic energy). Hence, the magnitude of the guiding-center
energy projection vector (\ref{eq:Delta_p}) is $|\vb{\Delta}^{p}|\equiv1$
up to second order in magnetic-field nonuniformity.

\subsubsection{Guiding-center Pitch-angle Projection}

The guiding-center pitch-angle projection vector (\ref{eq:Delta_xi0})
is expressed as 
\begin{eqnarray}
\vb{\Delta}^{\xi_{0}} & = & \frac{\sqrt{1-\xi_{0}^{2}}}{p\;\xi_{0}}\left[\frac{\xi}{\sqrt{\Psi}}\;\wh{\xi}\;+\;\epsilon\;\sqrt{1-\xi_{0}^{2}}\left(\vb{\Delta}_{1}^{p}\;-\;\frac{(\partial\vb{\rho}_{1}/\partial\zeta_{{\rm g}})}{\rho_{0}\;(1-\xi_{0}^{2})}\;-\;\frac{\rho}{2}\;\bhat\btimes\nabla\ln B_{0}\right)\right]\nonumber \\
 & \equiv & \frac{\sqrt{1-\xi_{0}^{2}}}{p\;\xi_{0}}\;\frac{\xi\;\wh{\xi}}{\sqrt{\Psi}}\;+\;\epsilon\;\vb{\Delta}_{1}^{\xi_{0}},
\label{eq:Deltaxi_exp}
\end{eqnarray}
 where $\vb{\Delta}_{1}^{\xi_{0}}$ denotes the first-order correction
of $\vb{\Delta}^{\xi_{0}}$. To lowest order in magnetic-field nonuniformity,
we find $|\vb{\Delta}^{\xi_{0}}|=\sqrt{1-\xi_{0}^{2}}\,\xi/(p\,\sqrt{\Psi}\;\xi_{0})$.

\subsection{Guiding-center Fokker-Planck Equation}

Now that the guiding-center Fokker-Planck components are expressed in terms of the invariant-space coordinates $I^{a}=(\ov{\psi},p,\xi_{0})$, we turn our attention to the guiding-center Fokker-Planck kinetic equation. When the original guiding-center coordinates $(\psi,\theta,\phi;{\cal E},\mu,
\zeta_{{\rm g}})$ are used in describing guiding-center dynamics in axisymmetric magnetic geometry (where we replace the guiding-center parallel velocity $v_{\|}$ with the guiding-center energy ${\cal E}$ and the gyroangle $\zeta_{{\rm g}}$ is an ignorable coordinate \cite{footnote_2}), the guiding-center
Vlasov evolution operator is ordered as \cite{RGL_83} 
\begin{equation}
\frac{d_{{\rm gc}}}{dt}\;\equiv\;\epsilon_{\tau}\;\pd{}{\tau}\;+\;\left(v_{\|}\,\frac{B^{\theta}}{B}\;+\;\epsilon_{{\rm B}}\,\dot{\theta}_{{\rm B}}\right)\pd{}{\theta}\;+\;\epsilon_{{\rm B}}\,\dot{\psi}_{{\rm B}}\;\pd{}{\psi},
\label{eq:dgc_dt}
\end{equation}
where we used $\partial/\partial\phi\equiv0$ (by definition of axisymmetry),
explicit time dependence is over long time scales (i.e., $\partial/\partial t\equiv\epsilon_{\tau}\,\partial/\partial\tau$
is ordered small), and the guiding-center dynamics on the poloidal plane is represented by Eq.~(\ref{eq:psitheta_dot}).

For each generic set of guiding-center invariants $(\ov{\psi}, {\cal E}, \mu)$, the guiding-center Fokker-Planck kinetic equation is expressed as
\begin{equation}
\epsilon_{\tau}\;\pd{F}{\tau}\;+\;\dot{\theta}\left(\pd{}{\theta}\;+\;\epsilon_{{\rm B}}\;\pd{\delta\psi}{\theta}\;\pd{}{\psi}\right)F\;=\;\epsilon_{\nu}\;{\cal C}_{{\rm gc}}[F],
\label{eq:gcFP_low}
\end{equation}
 where the magnetic-flux drift motion 
\begin{equation}
\dot{\psi} \;\equiv\; \pd{\delta\psi}{\theta}\;\dot{\theta}
\label{eq:psi_dot_B}
\end{equation}
is expressed in terms of the bounce radius (\ref{eq:psi_theta}) when
it is projected onto the guiding-center orbit ${\cal O}$ associated with the guiding-center invariants $(\ov{\psi}, {\cal E}, \mu)$. While the
topology of these orbits can be rather complicated \cite{Rome_Peng,Review_orbits,Hsu_Sigmar,Egedal,CCO}, our discussion will remain as general as possible concerning the nature of the axisymmetric magnetic geometry. Whether a guiding-center orbit ${\cal O}$ corresponds to a trapped-particle orbit or a passing-particle orbit, however, the motion is periodic in the poloidal plane $(\psi,\theta)$ and a generic orbital period in the poloidal plane 
\begin{equation}
\tau_{{\cal O}}\;\equiv\;\oint_{{\cal O}}\;\frac{d\theta}{\dot{\theta}}
\label{eq:tau_orbit_def}
\end{equation}
is defined as a closed-loop integral along the drift-orbit curve ${\cal O}$
parameterized by the poloidal angle $\theta$ (at constant invariants
$\ov{\psi}$, ${\cal E}$, and $\mu$). Note that the definition (\ref{eq:tau_orbit_def}) of the orbital period can also be given in terms of 
$d\psi/\dot{\psi}$ \cite{Rome_Peng} which, when evaluated along the guiding-center orbit ${\cal O}$, yields 
$(d\psi/\dot{\psi})_{\cal O} \equiv d\theta\,(\partial\delta\psi/\partial\theta)/\dot{\psi} = d\theta/\dot{\theta}$ upon using Eq.~(\ref{eq:psi_dot_B}).

\section{\label{sec:bounce_average}Orbit-averaged Guiding-center Fokker-Planck
Equation in Low-Collisionality Regime}

\subsection{Orbit-Averaging Operator in Axisymmetric Magnetic Geometry and Low
Collisionality Regime}

In the neoclassical transport regime $\epsilon_{\nu}=\nu\tau_{{\cal O}}\ll1$,
the guiding-center Fokker-Planck kinetic equation (\ref{eq:gcFP_low})
reduces to leading order to $\dot{\theta}\;\partial F/\partial\theta=0$,
which states that $F\;\equiv\;\ov{F}(\ov{\psi},{\cal E},\mu;\tau)$
is independent of the poloidal angle $\theta$. The physical picture
is that guiding centers orbits undergo many poloidal cycles before
being perturbed by collisions. 

We now introduce the \textit{orbit}-averaging operation 
\begin{equation}
\langle\cdots\rangle_{{\cal O}}\;\equiv\;\frac{1}{\tau_{{\cal O}}}\;\oint_{{\cal O}}\;(\cdots)\;\frac{d\theta}{\dot{\theta}},
\label{eq:average_orbit}
\end{equation}
where the orbital period $\tau_{\cal O}$ is defined in Eq.~(\ref{eq:tau_orbit_def}). By orbit-averaging the guiding-center Fokker-Planck kinetic
equation (\ref{eq:gcFP_low}) in the low-collisionality approximation,
we finally obtain the orbit-averaged guiding-center Fokker-Planck
kinetic equation 
\begin{equation}
\epsilon_{\tau}\;\pd{\ov{F}}{\tau}\;=\;\epsilon_{\nu}\;\left\langle {\cal C}_{{\rm gc}}[\ov{F}]\right\rangle _{{\cal O}},
\label{eq:OgcFP_eq}
\end{equation}
 which describes the collisional time evolution of the orbit-averaged
guiding-center distribution 
\begin{equation}
\langle F\rangle_{{\cal O}}\;=\;\ov{F}(\ov{\psi},{\cal E},\mu;\tau),
\label{eq:Fgc_O}
\end{equation}
and the guiding-center Fokker-Planck operator is given by Eq.~(\ref{eq:gcFP_def}).

Lastly, we note that the relation between the magnetic flux $\psi$
and its orbit-averaged value $\langle\psi\rangle_{{\cal O}}$ (the
{}``drift-center'' position) is expressed as \cite{Bernstein_Molvig}
\begin{equation}
\psi\;\equiv\;\langle\psi\rangle_{{\cal O}}\;+\;\left(\delta\psi\;-\frac{}{}\langle\delta\psi\rangle_{{\cal O}}\right),
\label{eq:psi_orbit}
\end{equation}
 where the drift-surface label is $\ov{\psi}\equiv\langle\psi\rangle_{{\cal O}}-\langle\delta\psi\rangle_{{\cal O}}$.
For trapped and passing particles, we therefore obtain the following
relations for the deviations from a magnetic-flux surface: 
\begin{equation}
\left(\begin{array}{c}
\psi\;-\;\langle\psi\rangle_{{\rm b}}\\
\\
\psi\;-\;\langle\psi\rangle_{{\rm t}}\end{array}\right)\;\equiv\;\left(\begin{array}{c}
\delta\psi\\
\\
\delta\psi\;-\;\langle\delta\psi\rangle_{{\rm t}}
\end{array}\right),
\label{eq:psi_bt}
\end{equation}
 where we used the fact that $\langle\delta\psi\rangle_{{\rm b}}\equiv0$
for trapped particles, while $\langle\delta\psi\rangle_{{\rm t}}\neq0$
for passing particles (since $\delta\psi$ does not change sign along
a passing-particle orbit). These relations show that, while the deviation
$\delta\psi$ for a trapped particle may be large, the deviation $(\delta\psi-\langle\delta\psi\rangle_{{\rm t}})$
for a passing particle is in general small.

\subsection{Approximate Orbit-Averaging Operator}

In this Section, an explicit expression for the orbit-averaged guiding-center
Fokker-Planck operator that appears on the right side of Eq.~(\ref{eq:OgcFP_eq})
is derived. In general, the orbit-averaging operation (\ref{eq:average_orbit})
must be computed along numerically calculated orbits. However, useful
analytical expressions of the orbit-averaged guiding-center Fokker-Planck
operator can be obtained when the orbit deviation $\delta\psi/|\nabla\psi|$
is small as compared to the local minor radius $r$. The difference
between the exact incremental time element $d\theta/\dot{\theta}$
and the approximate time element $ds/v_{\|}$ is of order $\epsilon_{{\rm B}}^{2}$
and can be neglected when compared to the corrections of order $\epsilon_{{\rm B}}$
in the Fokker-planck operator. The \textit{orbital} period (\ref{eq:tau_orbit_def}) thus becomes 
\begin{equation}
\tau_{{\cal O}}(\ov{\psi},{\cal E},\mu)\;\simeq\;\oint_{{\cal O}}\frac{ds}{v_{\|}}\;=\;\frac{1}{v}\;\oint_{{\cal O}}\;
\frac{d\theta}{\xi(\wt{\psi};{\cal E},\mu)}\;\frac{B(\wt{\psi},\theta)}{B^{\theta}(\wt{\psi},\theta)}.
\label{eq:taupar_def}
\end{equation}
The orbit topology is explicitly taken into account in Eq.~(\ref{eq:taupar_def}) through the dependence of the integrand $B/(\xi B^{\theta})$ on the bounce-radius $\delta\psi(\theta,\sigma;\ov{\psi},{\cal E},\mu)\equiv\wt{\psi}-\ov{\psi}$ defined in Eq.~(\ref{eq:psi_theta}), where 
$\ov{\psi}=\psi_{{\rm b}}$ (for trapped particles) or $\psi_{{\rm t}}$ (for passing particles). Note that the orbital period (\ref{eq:taupar_def}) is exact for stagnation orbits (where $\dot{\theta} = 0$ yields $\dot{\theta}_{\rm B} = -\,v_{\|}\,B^{\theta}/B$).

In the definition of the orbital period (\ref{eq:taupar_def}), it is convenient to extract the magnetic-geometric factor 
\begin{equation}
\Lambda(\ov{\psi},p,\xi_{0})\;\equiv\;\int_{-\pi}^{\pi}\; d\theta\;\frac{B(\wt{\psi},\theta)}{B^{\theta}(\wt{\psi},\theta)},
\label{eq:Lambda_def}
\end{equation}
 which defines the length of an orbit on the drift-surface labeled
by $\ov{\psi}$. In the standard axisymmetric tokamak ordering, we
find $\Lambda\simeq2\pi\, q(\ov{\psi})\, R_{0}$, where $R_{0}$ denotes
the major radius of the magnetic axis. Next, we define the normalized orbital period 
\begin{equation}
\lambda(\ov{\psi},p,\xi_{0})\;\equiv\;\frac{1}{\Lambda}\;\oint_{{\cal O}}\; d\theta\;\frac{\xi_{0}\, B}{\xi\, B^{\theta}}\;=\;\left(\frac{v\,\xi_{0}}{\Lambda}\right)\;\tau_{{\cal O}},
\label{eq:lambda_def}
\end{equation}
 which satisfies the condition $\lambda(\psi_{{\rm t}},p,|\xi_{0}|=1)\equiv1$
for completely-passing particles (i.e., $|\xi|=1$ and $\mu=0$ so
that $v_{\|}=v$ is a constant of motion). Note that the orbital period
(\ref{eq:taupar_def}) {[}or normalized orbital period (\ref{eq:lambda_def}){]}
becomes infinite on the boundary that separates trapped-particle orbits
and passing-particle orbits (i.e., pinch orbits). Hence, the low-collisionality approximation
$(\partial F/\partial\theta=0$) does not technically hold very close
to the trapped-passing boundary, where the bounce and transit periods
become much larger than the characteristic collisional time scale
$\nu^{-1}$. However, the fraction of particles with $\lambda\gg1$
is very small since the corresponding singularity is integrable. Indeed,
by introducing $\xi_{0{\rm b}}$ such that $1-\Psi\,(1-\xi_{0{\rm b}}^{2})=0$, it can be shown that 
\[
\int_{0}^{1}\;\lambda(\ov{\psi},p,\xi_{0})\, d\xi_{0}\;=\;\frac{1}{\Lambda}\;\int_{-\pi}^{\pi}\;\frac{B}{B^{\theta}}\; d\theta\;\int_{\xi_{0{\rm b}}}^{1}\;\frac{\xi_{0}\; d\xi_{0}}{\sqrt{1\;-\;\Psi\;\left(1-\xi_{0}^{2}\right)}}\;=\;\frac{1}{\Lambda}\;\int_{-\pi}^{\pi}\;\frac{B_{0}}{B^{\theta}}\; d\theta,
\]
 which is a geometrical factor of order unity.

Using the Jacobian (\ref{eq:Jac_def}), the normalized bounce period
(\ref{eq:lambda_def}) is now expressed as 
\begin{equation}
\lambda(\ov{\psi},p,\xi_{0})\;=\;\frac{B_{0}}{\Lambda p^{2}}\;\oint_{{\cal O}}\;{\cal J}\; d\theta\;\equiv\;2\pi\,{\cal J}_{\mathcal{O}}\;\frac{B_{0}}{\Lambda p^{2}},
\label{eq:lambda_Jac}
\end{equation}
 so that the orbital period (\ref{eq:taupar_def}) becomes 
\begin{equation}
\tau_{{\cal O}}\;=\;\frac{2\pi\,{\cal J}_{\mathcal{O}}\, B_{0}}{vp^{2}\,\xi_{0}}.
\label{eq:tau_J}
\end{equation}
The orbit-averaging operation (\ref{eq:average_orbit}) becomes 
\begin{equation}
\langle\cdots\rangle_{{\cal O}}(\ov{\psi},p,\xi_{0})\;\equiv\;\frac{1}{\tau_{{\cal O}}}\;\oint_{{\cal O}}\;(\cdots)\;\frac{ds}{v_{\|}}\;=\;\frac{1}{\tau_{{\cal O}}}\;\oint_{{\cal O}}\;(\cdots)\;\frac{B\; d\theta}{v\,\xi\, B^{\theta}}\;=\;\frac{1}{{\cal J}_{\mathcal{O}}}\;\oint_{{\cal O}}\,(\cdots)\,{\cal J}\;\frac{d\theta}{2\pi},
\label{eq:A_orbit}
\end{equation}
For trapped-particle orbits, the orbit-average (\ref{eq:A_orbit})
yields the explicit formula for bounce-averaging operation 
\begin{equation}
\langle\cdots\rangle_{\sf b}(\psi_{{\rm b}},p,\xi_{0})\;\equiv\;\frac{1}{\tau_{{\rm b}}}\;\sum_{\sigma}\;\int_{\theta_{{\rm b}}^{-}}^{\theta_{{\rm b}}^{+}}\;(\cdots)\;\frac{d\theta}{v|\xi|}\;\frac{B}{B^{\theta}}\;=\;\frac{1}{{\cal J}_{\mathcal{O}}}\;\left[\frac{1}{2}\,\sum_{\sigma}\right]\;\int_{\theta_{{\rm b}}^{-}}^{\theta_{{\rm b}}^{+}}\;(\cdots)\;{\cal J}\;\frac{d\theta}{2\pi}.
\label{eq:A_bounce}
\end{equation}
 Here, the summation is over the sign $\sigma$ of $\xi$ and, therefore,
the symbol $\frac{1}{2}\,\sum_{\sigma}$ takes the average (for trapped
particles) between values of the integrand for $\xi>0$ and values
of the integrand for $\xi<0$. For passing-particle orbits, on the
other hand, the orbit-average (\ref{eq:A_orbit}) yields the explicit
formula for transit-averaging operation 
\begin{equation}
\langle\cdots\rangle_{\sf t}(\psi_{{\rm t}},p,\xi_{0})\;\equiv\;\frac{1}{{\cal J}_{\mathcal{O}}}\;\int_{0}^{2\pi}\;(\cdots)\;{\cal J}\;
\frac{d\theta}{2\pi}.
\label{eq:A_transit}
\end{equation}

\subsection{\label{subsec:bgcFP}Orbit-averaged Guiding-center Fokker-Planck Operator}

In the expression for the guiding-center Fokker-Planck operator (\ref{eq:gcFP_def}),
the guiding-center function $\ov{F}$ is independent of the angle coordinates
$(\theta,\phi,\zeta_{{\rm g}})$, in the limit of weak collisionality
and under the assumptions of axisymmetry and gyroangle invariance.
Hence, only derivatives of the guiding-center distribution $\ov{F}$ with
respect to the guiding-center invariants $I^{a}$ will remain in Eq.~(\ref{eq:gcFP_def}).
The coefficients $({\cal K}_{{\rm gc}}^{a},{\cal D}_{{\rm gc}}^{ab})$
and the Jacobian ${\cal J}$ in the guiding-center Fokker-Planck operator
(\ref{eq:gcFP_def}), on the other hand, still depend on the poloidal
angle $\theta$. We therefore need to orbit-average the guiding-center
Fokker-Planck operator (\ref{eq:gcFP_def}) according to the orbit-averaging
procedure (\ref{eq:A_orbit}).

The orbit-averaged guiding-center Fokker-Planck operator is expressed as 
\begin{equation}
\left\langle {\cal C}_{{\rm gc}}[\ov{F}]\right\rangle _{{\cal O}}\;\equiv\;-\;\frac{1}{{\cal J}_{\mathcal{O}}}\;\pd{}{I^{a}}
\left[\;{\cal J}_{\mathcal{O}}\left(\left\langle {\cal K}_{{\rm gc}}^{a}\right\rangle _{{\cal O}}\; \ov{F} \;-\; \left\langle {\cal D}_{{\rm gc}}^{ab}\right\rangle _{{\cal O}}\;\pd{\ov{F}}{I^{b}}\right)\;\right],
\label{eq:gcFP_bounce}
\end{equation}
 where the averaged Jacobian ${\cal J}_{\mathcal{O}}$ now becomes
the Jacobian for the orbit-averaged guiding-center Fokker-Planck operator
(\ref{eq:gcFP_bounce}). Orbit-averaged guiding-center Fokker-Planck
operator that treat friction and diffusion in three-dimensional space
have been derived previously \cite{ZOC_93}. While these previous
operators retained only the lowest-order terms in magnetic-field nonuniformity,
the orbit-averaged guiding-center Fokker-Planck operator (\ref{eq:gcFP_bounce})
retains first-order corrections as well, which allows a realistic
magnetic geometry to be considered.

The orbit-averaged guiding center Fokker-Planck (FP) operator is widely
used in the physics of magnetized plasmas (earth magnetosphere \cite{Brizard_rql_04},
thermonuclear fusion \cite{FP_review,kar86}, ...). For practical
applications (like studying fast particle dynamics generated by rf
waves or a constant electric field \cite{Peysson_08b}), analytical
expressions often correspond to an oversimplified description of the
problem of interest, and therefore a more realistic approach requires
full numerical calculations.

For computational purposes, the orbit-averaged guiding-center Fokker-Planck
operator (\ref{eq:gcFP_bounce}) is recast in a flux-conservative form 
\begin{equation}
\left\langle {\cal C}_{{\rm gc}}[\ov{F}]\right\rangle _{{\cal O}}\;\equiv\;\frac{1}{{\cal J}_{\mathcal{O}}}\frac{\partial}{\partial\ov{\psi}}\left({\cal J}_{\mathcal{O}}|\nabla\ov{\psi}|_{0}\; S_{\sf L}^{\ov{\psi}}\right)\;+\;\frac{1}{p^{2}}\frac{\partial}{\partial p}\left(p^{2}\frac{}{}S_{\sf L}^{p}\right)-\frac{1}{\lambda p}\frac{\partial}{\partial\xi_{0}}\left(\lambda\sqrt{1-\xi_{0}^{2}}\; S_{\sf L}^{\xi_{0}}\right),
\label{eq:BAVGCFPE2}
\end{equation}
 such that the usual two-grid discretization technique may be applied
for the finite-difference method \cite{PD_09a}. Particle conservation
is therefore naturally satisfied numerically up to second order in
the truncation error. In Eq.~(\ref{eq:BAVGCFPE2}), $S_{\sf L}^{\ov{\psi}}$
describes particle flux across magnetic flux surfaces, while $S_{\sf L}^{p}$
and $S_{\sf L}^{\xi_{0}}$ account for momentum and pitch-angle dynamics
respectively. From Eq.~(\ref{eq:gcFP_bounce}), we have 
\begin{equation}
\left(\begin{array}{l}
S_{\sf L}^{\ov{\psi}}\\
\\
S_{\sf L}^{p}\\
\\
S_{\sf L}^{\xi_{0}}\end{array}\right)\;\equiv\;\left(\begin{array}{l}
K_{\sf L}^{\ov{\psi}}\\
\\
K_{\sf L}^{p}\\
\\
K_{\sf L}^{\xi_{0}}\end{array}\right)\;\ov{F}\;-\;\left(\begin{array}{lll}
D_{\sf L}^{\ov{\psi}\ov{\psi}} & D_{\sf L}^{\ov{\psi}p} & D_{\sf L}^{\ov{\psi}\xi}\\
\\
D_{\sf L}^{p\ov{\psi}} & D_{\sf L}^{pp} & D_{\sf L}^{p\xi}\\
\\
D_{\sf L}^{\xi_{0}\ov{\psi}} & D_{\sf L}^{\xi_{0}p} & D_{\sf L}^{\xi_{0}\xi_{0}}\end{array}\right)\;\left(\begin{array}{c}
|\nabla\ov{\psi}|_{0}\;\partial/\partial\ov{\psi}\\
\\
\partial/\partial p\\
\\
-\, p^{-1}\sqrt{1-\xi_{0}^{2}}\;\partial/\partial\xi_{0}\end{array}\right)\;\ov{F},
\label{eq:fluxdef}
\end{equation}
 where the friction coefficients coefficients $K_{\sf L}^{a}$ and
symmetric diffusion coefficients $D_{\sf L}^{ab}=D_{\sf L}^{ba}$
may be expressed in terms of guiding-center friction and diffusion
components (\ref{eq:Kgc}) and (\ref{eq:Dgc}) according to the general
relations 
\begin{eqnarray}
K_{\sf L}^{\ov{\psi}} & = & \frac{1}{|\nabla\ov{\psi}|_{0}}\left\langle \mathcal{K}_{{\rm gc}}^{\ov{\psi}}\right\rangle _{{\cal O}}\\
K_{\sf L}^{p} & = & \left\langle \mathcal{K}_{{\rm gc}}^{p}\right\rangle _{{\cal O}}\\
K_{\sf L}^{\xi_{0}} & = & -\frac{p}{\sqrt{1-\xi_{0}^{2}}}\left[\frac{1-\xi_{0}^{2}}{2\xi_{0}}\left\langle \frac{1}{\Psi}
\nabla\ln\Psi\bdot\mathcal{K}_{{\rm gc}}^{{\bf X}}\right\rangle _{{\cal O}}+\left\langle \frac{1}{\Psi}\frac{\xi}{\xi_{0}}
\mathcal{K}_{{\rm gc}}^{\xi_{0}}\right\rangle _{{\cal O}}\right]
\end{eqnarray}
 and 
\begin{eqnarray}
D_{\sf L}^{\ov{\psi}\ov{\psi}} & = & \frac{1}{|\nabla\ov{\psi}|_{0}^{2}}\left\langle \mathcal{D}_{{\rm gc}}^{\ov{\psi}\ov{\psi}}\right\rangle _{{\cal O}}\\
D_{\sf L}^{\ov{\psi}p} & = & \frac{1}{|\nabla\ov{\psi}|_{0}}\left\langle \mathcal{D}_{{\rm gc}}^{\ov{\psi}p}\right\rangle _{{\cal O}}\\
D_{\sf L}^{\ov{\psi}\xi_{0}} & = & -\frac{p}{\sqrt{1-\xi_{0}^{2}}}\frac{1}{|\nabla\ov{\psi}|_{0}}\left[\frac{1-\xi_{0}^{2}}{2\xi_{0}}\left\langle \frac{1}{\Psi}\nabla\ov{\psi}\bdot\mathcal{D}_{{\rm gc}}^{{\bf X}{\bf X}}\bdot\nabla\ln\Psi\right\rangle _{{\cal O}}+\left\langle \frac{1}{\Psi}\frac{\xi}{\xi_{0}}\;\mathcal{D}_{{\rm gc}}^{\ov{\psi}\xi_{0}}\right\rangle _{{\cal O}}\right]\\
D_{\sf L}^{pp} & = & \left\langle \mathcal{D}_{{\rm gc}}^{pp}\right\rangle _{{\cal O}}\\
D_{\sf L}^{p\xi_{0}} & = & -\frac{p}{\sqrt{1-\xi_{0}^{2}}}\left[\left\langle \frac{1}{\Psi}\frac{\xi}{\xi_{0}}\mathcal{D}_{{\rm gc}}^{p\xi_{0}}\right\rangle _{{\cal O}}+\frac{1-\xi_{0}^{2}}{2\xi_{0}}\left\langle \frac{1}{\Psi}\mathcal{D}_{{\rm gc}}^{p{\bf X}}\bdot\nabla\ln\Psi\right\rangle _{{\cal O}}\right]\\
D_{\sf L}^{\xi_{0}\xi_{0}} & = & \frac{p^{2}}{1-\xi_{0}^{2}}\left[\left\langle \frac{1}{\Psi^{2}}\frac{\xi^{2}}{\xi_{0}^{2}}\mathcal{D}_{{\rm gc}}^{\xi_{0}\xi_{0}}\right\rangle _{{\cal O}}+\frac{1-\xi_{0}^{2}}{\xi_{0}}\left\langle \frac{1}{\Psi^{2}}\frac{\xi}{\xi_{0}}\mathcal{D}_{{\rm gc}}^{\xi_{0}{\bf X}}\bdot\nabla\ln\Psi\right\rangle _{{\cal O}}\right].
\end{eqnarray}

The numerical implementation of the orbit-averaged guiding-center
Fokker-Planck operator (\ref{eq:BAVGCFPE2}) is described in a forthcoming
paper \cite{PD_09a} using a novel 3-D scheme with fully-implicit
time evolution. It is incorporated in the code \textsf{LUKE} \cite{DP_09},
which was initially developed for fast electron physics, and will therefore
extend its range of applicability to multispecies physics and transport
processes.

\section{\label{sec:bgcFP}Bounce-center Fokker-Planck Operator}

The next step in the derivation of a reduced Fokker-Planck operator
is to proceed with the construction of a \textit{bounce-center} Fokker-Planck
operator 
\begin{equation}
{\cal C}_{{\rm bc}}[\wh{F}]\;\equiv\;\left\langle {\sf T}_{{\rm bc}}^{-1}{\cal C}_{{\rm gc}}\left[{\sf T}_{{\rm bc}}\frac{}{}\wh{F}\right]
\right\rangle _{{\rm b}},
\label{eq:C_bc_def}
\end{equation}
 where ${\sf T}_{{\rm bc}}^{-1}$ and ${\sf T}_{{\rm bc}}$ are the
push-forward and pull-back operators associated with the bounce-center
phase-space transformation \cite{Cary_Brizard} (and references therein),
$\wh{F}\equiv{\sf T}_{{\rm bc}}^{-1}F$ denotes the bounce-center
distribution (which is independent of the bounce angle $\zeta_{{\rm b}}$),
and $\langle\;\rangle_{{\rm b}}$ denotes averaging with respect to
$\zeta_{{\rm b}}$. In the bounce-center Fokker-Planck operator (\ref{eq:C_bc_def}),
finite-orbit effects will explicitly be taken into account.

To see how these finite-orbit effects might arise in the bounce-center
Fokker-Planck operator (\ref{eq:C_bc_def}), we first express the
bounce-center push-forward and pull-back operators as ${\sf T}_{{\rm bc}}^{\pm}\equiv\exp(\pm\,\epsilon\,{\sf G}_{1}\cdot\exd)$,
where ${\sf G}_{1}\cdot\exd\equiv G_{1}^{a}\,\partial_{a}$ is defined
in terms of the components $G_{1}^{a}$ of the first-order generating
vector field for the bounce-center transformation and $\epsilon\equiv\epsilon_{{\rm d}}=\tau_{{\rm b}}/\tau_{{\rm d}}\ll1$
denotes its ordering parameter. Next, we expand the transformed operator
\begin{equation}
\wh{\cal J}\;{\sf T}_{{\rm bc}}^{-1}{\cal C}_{{\rm gc}}\left[{\sf T}_{{\rm bc}}\frac{}{}\wh{F}\right]\;=\;\wh{\cal J}\left(e^{-\,\epsilon\,{\sf G}_{1}\cdot\exd}\;{\cal C}_{{\rm gc}}\left[e^{\epsilon\,{\sf G}_{1}\cdot\exd}\;\wh{F}\right]\right)
\label{eq:G_bgc_exact}
\end{equation}
 in powers of $\epsilon$, where the bounce-center Jacobian $\wh{\cal J}$
is defined in terms of the guiding-center Jacobian ${\cal J}$ as
$\wh{\cal J}\equiv{\cal J}-\epsilon\;\partial_{a}(G_{1}^{a}\,{\cal J})+\cdots$.
By keeping terms up to second order in $\epsilon$, and rearranging
terms, we obtain 
\begin{equation}
\left(\wh{\cal J}\;+\;\epsilon\;\partial_{a}(G_{1}^{a}\wh{\cal J})\;+\frac{}{}\cdots\right){\cal C}_{{\rm gc}}\left[\wh{F}\;+\;\epsilon\; G_{1}^{b}\,\wh{F}+\cdots\frac{}{}\right]\;-\;\epsilon\;\partial_{a}\left(\wh{\cal J}\, G_{1}^{a}\;{\cal C}_{{\rm gc}}\left[\wh{F}\;+\;\epsilon\; G_{1}^{b}\,\partial_{b}\wh{F}+\cdots\frac{}{}\right]\right)\;+\;\cdots,
\label{eq:C_bgc_quad}
\end{equation}
 where the first term is simply the original guiding-center Fokker-Planck
operator multiplied by the guiding-center Jacobian ${\cal J}\equiv\wh{\cal J}+\epsilon\;\partial_{a}(G_{1}^{a}\wh{\cal J})+\cdots$
(i.e., the first term is an exact divergence), while the second term
is also an exact divergence in bounce-center phase space. The total
bounce-center Fokker-Planck operator is therefore guaranteed to retain
its phase-space divergence form.

In future work, it will be our purpose to show that finite-orbit effects
will appear in the bounce-center Fokker-Planck operator (\ref{eq:C_bc_def})
in the form of the second-order term \[
\frac{\epsilon^{2}}{\wh{\cal J}}\;\pd{}{\wh{\psi}}\left\langle \wh{\cal J}\, G_{1}^{\psi}\;{\cal C}_{{\rm gc}}\left[G_{1}^{\psi}\,\pd{\wh{F}}{\wh{\psi}}\right]\right\rangle _{{\rm b}},\]
 where $\wh{\psi}$ denotes the bounce-center position and $G_{1}^{\psi}$
denotes the corresponding component of the bounce-angle-dependent
bounce-radius. In the present work, $\langle\psi\rangle_{{\cal O}}\equiv\ov{\psi}$
plays the role of the bounce-center position $\wh{\psi}$ for a trapped-particle
orbit ($\delta\psi$ plays the role of $G_{1}^{\psi}$), while the
nonlocal bounce-action $J_{{\rm b}}$ (canonically conjugate to the
bounce angle $\zeta_{{\rm b}}$) is represented by the minimum-B pitch-angle
coordinate $\xi_{0}$.

\section{\label{sec:sum}Summary}

The present work presents the derivation of a general guiding-center Fokker-Planck equation in axisymmetric magnetic geometry (\ref{eq:gcFP_low}) that allows first-order corrections in magnetic-field nonuniformity to be retained. The Fokker-Planck operator (\ref{eq:gcFP_def}) is obtained from a guiding-center transformation using the set of invariants $(\ov{\psi},p,\xi_{0})$. Next, the orbit-averaging procedure is introduced and an explicit
Fokker-Planck operator (\ref{eq:gcFP_bounce}) is derived in the low-collisionality limit. This
operator is also expressed in a conservative form (\ref{eq:BAVGCFPE2}) that is best suited for numerical application.  

In the continuation of this paper (based on work briefly outlined in Sec.~\ref{sec:bgcFP}), future theoretical work will consider finite-orbit effects in a different way by deriving a general bounce-center Fokker-Planck operator using Lie-transform methods.

\acknowledgements

One of us (AJB) would like to thank the Institut de la Recherche sur
la Fusion par confinement Magn\'{e}tique at CEA Cadarache for their kind
hospitality. This work, supported by the European Communities under
the contract of Association between EURATOM and CEA, was carried out
within the framework of the European Fusion Development Agreement.
The views and opinions expressed herein do not necessarily reflect
those of the European Commission.


\begin{thebibliography}{23}
\bibitem{footnote_1} We note that the safety factor $q$ is equal
to zero for an axisymmetric dipole magnetic field (since $B_{\phi}=B^{\phi}/|\nabla\phi|^{2}\equiv0$).
\bibitem{footnote_2} It is important to note here that the guiding-center
distribution is independent of the gyroangle $\zeta_{{\rm g}}$ in
the absence of collisions only if the full guiding-center phase-space
transformation (with higher-order corrections associated with magnetic-field
nonuniformity \cite{RGL_83}) is considered.
\end{thebibliography}
\end{document}